\documentclass[
    a4paper,
    doc,
    natbib
]{apa6}

\usepackage[british]{babel}
\usepackage[utf8]{inputenc}
\usepackage{epstopdf}
\usepackage{csquotes}
\usepackage[hidelinks]{hyperref}
\usepackage{multirow}
\usepackage{multicol}
\usepackage{pdflscape}

\newcommand{\citeowner}[1]{\citeauthor{#1}'s (\citeyear{#1})}

\title{Identifying Relationships Between Cognitive Processes Across Tasks, Contexts, and Time}
\shorttitle{Estimating Parameter Correlations}

\author{Laura Wall$^1$, David Gunawan$^2$, Scott D. Brown$^1$, Minh-Ngoc Tran$^3$, Robert Kohn$^4$, Guy E. Hawkins$^1$}

\affiliation{1: School of Psychology, University of Newcastle, Australia \\
 2: School of Mathematics and Applied Statistics, University of Wollongong \\
 3: Discipline of Business Analytics, University of Sydney Business School \\
 4: Australian School of Business, University of New South Wales, Sydney, Australia}
\authornote{This research was partially supported by the Australian Research Council (ARC) Discovery Project scheme (DP180102195, DP180103613). Hawkins was partially supported by an ARC Discovery Early Career Researcher Award (DECRA; DE170100177).}

\abstract{
It is commonly assumed that a specific testing occasion (task, design, procedure, etc.) provides insights that generalise beyond that occasion. This assumption is infrequently carefully tested in data. We develop a statistically principled method to directly estimate the correlation between latent components of cognitive processing across tasks, contexts, and time. This method simultaneously estimates individual-participant parameters of a cognitive model at each testing occasion, group-level parameters representing across-participant parameter averages and variances, and across-task correlations. The approach provides a natural way to ``borrow'' strength across testing occasions, which can increase the precision of parameter estimates across all testing occasions. Two example applications demonstrate that the method is practical in standard designs. The examples, and a simulation study, also provide evidence about the reliability and validity of parameter estimates from the linear ballistic accumulator model. We conclude by highlighting the potential of the parameter-correlation method to provide an ``assumption-light'' tool for estimating the relatedness of cognitive processes across tasks, contexts, and time.}


\begin{document}
\maketitle

\newpage
\section{Introduction}

Evidence accumulation models of simple decisions, such as the linear ballistic accumulator \citep[LBA;][]{brown2008simplest} and the diffusion model \citep{ratcliff1998modeling}, began as theoretical tools to understand the cognitive processes of simple decision making. However, they are now increasingly used as psychometric tools in clinical and applied research. For example, there is extensive research using the diffusion and LBA models showing that older adults perform slower on simple cognitive tasks mostly because of changes in the speed with which motor response actions are executed, and not due to decreased processing speed, as was traditionally theorised \citep{ratcliff2004analysis,ratcliff2007application,ratcliff2006aging,forstmann2011speed-accuracy}. Other investigations have addressed questions about clinical disorders, for example finding differences in decision making processes for people with anxiety \citep{white2010anxiety}, depression \citep{ho2014functional}, schizophrenia \citep{heathcote2015decision,matzke2017failures}, and ADHD \citep{weigard2014diffusion}. 

In applied investigations using evidence accumulation models, researchers typically do not emphasise choices about the particular decision making task that is used. The task is usually chosen to be amenable to modelling, allowing many decisions in a session, with clearly-timed events within each one, and to have some validity as a measure of the cognitive process under investigation; e.g., a flanker task to measure attention, or a stopping task to measure inhibitory control. Despite the limitations of every decision task, investigators presumably intend their inferences to generalise beyond the chosen task. For example, \cite{ho2014functional} concluded that people with depression exhibit poorer perceptual sensitivity compared with a control group. This conclusion was based on the analysis of parameters estimated using data from a gender discrimination task. \citeauthor{ho2014functional} assumed that parameters estimated from other perceptual decision-making tasks would lead to similar results, for the same sample of participants. The more general assumption here is that there is some consistency in the parameter estimates across tasks for individuals. 

Given the extensive use of evidence accumulation models as measurement tools \citep{ratcliff2016diffusion}, there has been some investigation of the psychometric properties of the models, and particularly of the reliability and validity of the estimated parameters. \cite{voss2004interpreting} tested the criterion validity of the diffusion model by manipulating aspects of the task which could be expected to selectively influence different model components: manipulating the difficulty of the decision stimuli selectively influenced parameters related to processing rate, manipulating the cautiousness of the decision makers selectively influenced parameters which balanced urgency vs. caution, etc. Literally dozens of experiments have confirmed that model parameters related to processing speed are reliably affected by changes in the difficulty of the decision itself -- motion coherence, visual contrast, etc. Other experiments have investigated changes between people rather than between conditions. \cite{ratcliff2010individual} investigated the known-groups validity of the diffusion model by showing that individuals with a higher IQ also produce higher drift rate estimates. Similar studies have shown expected differences in diffusion and LBA parameters for people with depression \citep{ho2014functional}, anxiety \citep{white2010anxiety}, ADHD \citep{weigard2014diffusion}, and schizophrenia \citep{heathcote2015decision,matzke2017bayesian}.

The psychometric reliability of parameter estimates has been less carefully investigated than validity. Using the 
diffusion model, \cite{lerche2017retest} examined correlations between parameters estimated from lexical decisions and from recognition memory for pictures. Subjects in that experiment participated in two different sessions, and \citeauthor{lerche2017retest} observed only weak correlations in parameters across tasks for data from the first session. Data from the second session, however, provided stronger correlations. \cite{ratcliff2010individual} used two similar tasks (lexical decision, and recognition memory for words) and observed reliable correlations in almost all parameters of the diffusion model. \cite{ratcliff2015numeracy} investigated numeracy using four different decision-making tasks. In that investigation, parameters of the diffusion model related to processing speed correlated across tasks, but the other model parameters did not. \cite{mueller2019individual} also used the diffusion model, and analysed data from an experiment in which one group of participants completed two tasks related to emotion perception: one task used word-based stimuli, the other used faces. \citeauthor{mueller2019individual} found that parameters of the diffusion model related to response style and non-decision time were more strongly correlated across tasks than drift-related parameters, on average. Similarly, \cite{hedge2019slow} found moderate-to-good correlations between response caution parameters of the diffusion model across flanker, Stroop, and random dot motion tasks.

Clearly, the properties of the decision making task influence parameter estimates -- this is sometimes expected and desired, such as when stimulus properties related to decision difficulty influence drift rate estimates. However, it is important to establish that there is some reliable correlation in parameter estimates across tasks, in order to support the assumption that results observed using one particular decision making task can generalise to other, related, decisions. 

We investigate correlations in latent cognitive processes across tasks, using the LBA model. An important theoretical contribution of our work is that we directly estimate between-task parameter correlations as part of the model. Previous investigations have always estimated parameters for different tasks independently, and then examined correlations in those estimated parameters afterwards. Instead, our approach involves estimating parameters for multiple tasks simultaneously, while also estimating the correlations between those parameters. This approach has important statistical and methodological benefits, as well as scientific advantages. Estimating parameters using data from multiple tasks allows for ``borrowing'' of information across the tasks, analogous to the borrowing that takes place between participants in a repeated measures design. This improves estimation precision, especially for tasks with few data per person, and opens up exciting new possibilities. For example, some data collection procedures have subjects participate in several different decision-making tasks, such as those in a psychological test battery. This approach naturally restricts the amount of data collected for each individual task, making cognitive modelling of those tasks difficult or impossible. However, modelling the tasks jointly, and estimating the correlation in parameters across tasks, allows for information from one task to inform parameter estimates for other tasks. As long as some consistency in parameter estimates can be expected across tasks, this approach can allow analysis of data not previously possible.

\section{Applications}

We apply our methods to data from two decision-making experiments: one first reported by \cite{forstmann2008striatum}, and a new experiment. \citeauthor{forstmann2008striatum}'s experiment had $n=19$ participants repeatedly judge the direction of motion of a cloud of moving dots. On some decisions, participants were encouraged to be very urgent (``speed emphasis''), on other decisions they were encouraged to be very careful (``accuracy emphasis''), and on still others to balance speed and accuracy (``neutral emphasis''). Each participant practised the task for more than an hour, in a regular lab environment, and then later also performed the task while in a magnetic resonance imaging (MRI) scanner. See p.17541 of the original article for full details of the method. 

\cite{vanmaanen2016impact} investigated differences in performance between decisions made in and out of the scanner, using the LBA model, and found differences in parameter estimates from the two sessions. Our interest here is in the parameter correlations between sessions. Except for the differences induced by the scanner environment, Forstmann et al.'s experiment provides an opportunity to examine the reliability of the model parameters. There are several possible reasons why parameters estimated in and out of the scanner may differ: sampling error from the finite number of trials per person; different effects of the scanner environment on different people; and actual changes in the latent cognitive processing of the participants across time. Our investigation uncovers what commonality remains in parameter estimates beyond these effects. 

The experiment reported by \cite{forstmann2008striatum} used an identical decision-making task in the two sessions. What changes between sessions is the environmental context (the MRI scanner vs. the lab) and also the amount of data. The out-of-scanner session, which came first, involved more than three times as much data per-person as the in-scanner session. The second data set we analyse had participants undertake three tasks. The three tasks were chosen to share some common elements, including the basic visual properties of the stimulus, but to differ in their cognitive demands. One task used visual search -- finding a feature conjunction amongst distractors that shared the same features in different combinations. The difficulty of the visual search task was manipulated by changing the number of distractor items. Every display always included a target item, and the participant's task was to subsequently report the location of a search-irrelevant feature on that target. 

Another task was identical to the visual search task, but with an added component of response inhibition. In the ``stop'' task, a random 25\% of trials were interrupted by a signal which instructed the participant to withhold their response. The stop-signal task has become important for understanding inhibitory control \citep{logan1984on}, but it is also not well-suited to cognitive modelling \citep{matzke2017bayesian,matzke2017failures}. Following their recommendations, we restricted our analyses to data from trials which were not interrupted by a stop signal -- we did not model the stopping process. The third task used the same visual stimuli, but tested participants' short term memory. This ``match'' task required participants to decide whether the stimulus array shown on one trial had the same set of stimuli (perhaps in different locations) as the stimulus array shown on the preceding trial. The match task is a variant of the ``$n$-back'' task, which is a widely used and very difficult memory test. We manipulated the difficulty of the match task by changing the number of items in the stimulus array. Appendix A gives full details of the experiment.

\section{Modelling Correlations in Latent Cognitive Processes Across Tasks}

\begin{figure}
\centering\includegraphics[width=0.9\linewidth]{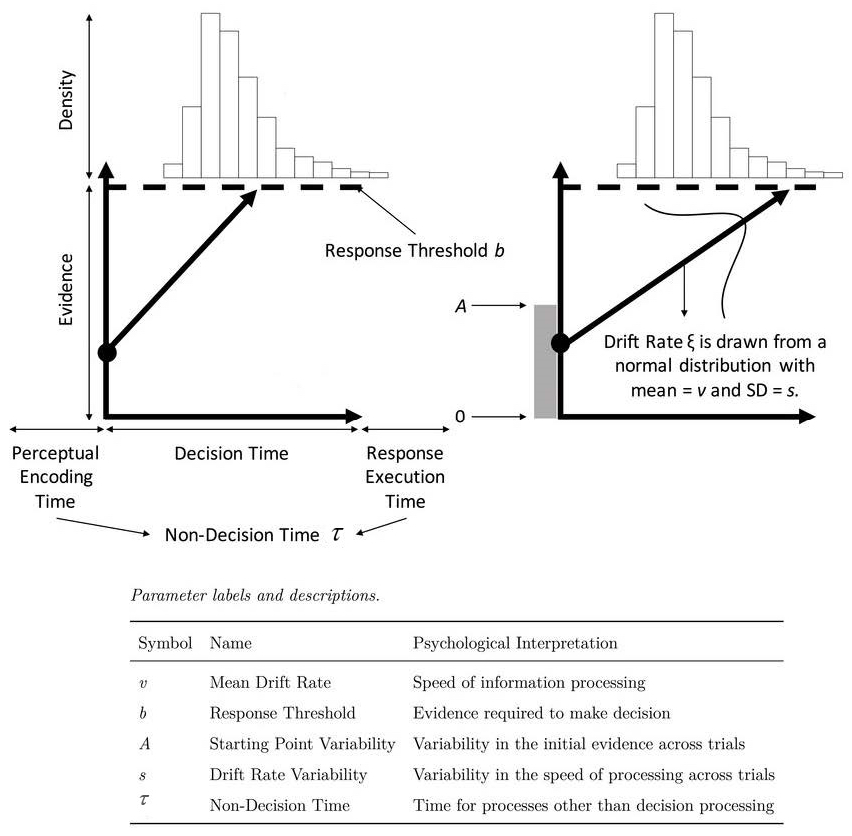}
\caption{The linear ballistic accumulator. On each trial, evidence for each response option begins randomly between $0$ and $A$; the upper value of the start point variability. The speed of the linear evidence accumulation is called the ``drift rate'', which is sampled from a normal distribution with mean $v$, unit standard deviation, truncated to positive values. Accumulation continues until a response threshold ($b$ units above $A$) is reached. The accumulator which reaches the threshold first (the left accumulator in this example) determines the response.}
\label{modelfig}
\end{figure}

We develop an approach to modelling across-task correlations in the latent processes by linking the parameters of evidence accumulation models of decision making across those tasks. Evidence accumulation models are named by their shared premise that when making a decision, evidence is accumulated for each choice alternative until a threshold amount is reached, which triggers a decision. For an LBA model of a two-choice decision, there are two accumulators, one corresponding to each response (see Figure \ref{modelfig}). The speed of evidence accumulation is called the ``drift rate'', and this randomly varies from decision-to-decision, reflecting changes in attention and internal states \citep{ratcliff1978theory}. In the LBA model, the distribution of drift rates is usually assumed to follow a normal distribution truncated to positive values, although other distributions are also possible \citep{terry2015generalising}. The mean of the drift rate distribution ($v$) is usually larger in an accumulator for a response alternative which matches the stimulus (a correct response) than one that does not, but on any particular trial, sample drift rates will be different. We assume a variance of $s^2=1$ for all drift rate distributions. The other source of random variability in the LBA model concerns the amount of evidence with which each accumulator begins. This ``starting point'' is randomly sampled for each accumulator and each decision, from a uniform distribution of width $A$. Evidence accumulation continues until the first accumulator reaches a threshold value $b$, which is larger than the maximum starting point. Threshold crossing triggers a response, which is delayed by some fixed constant $\tau$, representing the time taken for processes outside of decision-making, such as stimulus encoding and the execution of the motor response.

Reflecting the reality of inter-twined cognitive systems, and like all cognitive models, the parameters of the LBA model are correlated. For example, increases in the decision threshold lead to slower and more variable predicted response times, and more errors. Similar (but not identical) predictions can also arise from decreased mean drift rates. Parameter correlations can cause estimation difficulty, for example requiring more sophisticated sampling or search algorithms \citep{turner2013method}. We build on a recent advance in the literature, by \cite{gunawan2018new}, which directly estimates the correlations between parameters in the prior, and improves statistical efficiency. \citeauthor{gunawan2018new}'s method first log-transforms the parameters (both group, and individual-level) of the LBA model, so that they have support on the entire real line. The method then assumes that the distribution of log-transformed parameters across participants is multivariate normal. The correlations implied by that multivariate normal distribution describes dependence between parameters. 

Our article extends the method of \cite{gunawan2018new} to model dependence between tasks. We extend the vector of parameters for each person to include parameters for two or more tasks, so that the correlation matrix has a block-wise structure in which the diagonal blocks address within-task parameter dependence and the off-diagonal blocks address dependence in parameters between tasks. These off-diagonal blocks answer the question posed above, measuring the extent to which parameters from different tasks align. The correlation matrix also allows for statistical ``borrowing'' of strength between tasks, due to the inferred relationships between the tasks. Data and code for both applications reported below are available at \url{osf.io/rf8nd}.

\section{Results}

\subsection{Application 1: Correlations in Latent Processes In and Out of the Scanner}
 
To model the decisions in each session, we followed the same LBA specification as used in the original article \citep{forstmann2008striatum} and confirmed subsequently by \cite{gunawan2018new}. We collapsed across left- and right-moving stimuli, forcing the same mean drift rate for the accumulator corresponding to a ``right'' response to a right-moving stimulus as for the accumulator corresponding to a ``left'' response to a left-moving stimulus; we denote this mean drift rate by $v^{\left(c\right)}$. Similarly, drift rates for the accumulators corresponding to the wrong direction of motion are constrained to be equal and denoted by $v^{\left(e\right)}$. Three different response thresholds were estimated, for the speed, neutral and accuracy conditions: $b^{\left(s\right)}$, $b^{\left(n\right)}$ and $b^{\left(a\right)}$, respectively. Two other parameters were also estimated: the time taken by non-decision process ($\tau$) and the width of the uniform distribution for start points in evidence accumulation ($A$). 

These assumptions required estimating seven parameters: $\left(A, v^{\left(c\right)}, v^{\left(e\right)}, b^{\left(s\right)}, b^{\left(n\right)}, b^{\left(a\right)}, \tau  \right)$. Different parameters were estimated for the in-scanner and out-of-scanner sessions. The full vector of 14 (log-transformed) parameters was estimated as a random effect for each participant, with a multivariate normal prior distribution assumed across participants. The prior for the mean vector of the multivariate normal distribution is another multivariate normal distribution with zero mean, whose covariance matrix is the identity matrix. For the prior on the covariance matrix of the group distribution, we followed the recommendations of \cite{huang2013simple} and used a random mixture of inverse Wishart distributions, with mixture weights according to an inverse Gaussian distribution, which leads to marginally non-informative (uniform) priors on all correlation coefficients, and half-$t$ distributed priors on the standard deviations. These settings, and all other sampling details, are identical to those reported by \cite{gunawan2018new}.

Since the data from this experiment were used to estimate the LBA model previously, several times, our article does not report the usual summaries of the model's goodness-of-fit; see Figures 6 and 7 of \cite{vanmaanen2016impact} for details. Our focus is on the estimated parameters. Table~\ref{tab:forstmanmu} shows the estimated parameters separately for the two sessions. Compared with the out-of-scanner session, when participants were tested in the MRI scanner, the group average parameters changed in ways consistent with those reported by \citeauthor{vanmaanen2016impact}. In the scanner, participants made more cautious decisions (higher thresholds, $b$, and larger start point variability, $A$), but there was little difference in drift rate or non-decision time parameters. 

\begin{table}[ht]
\centering
\caption{Mean (and SD) of the estimated marginal posterior distributions for the LBA mean parameters, using data from Forstmann et al. (2008); see text for details. For ease of interpretation, these parameters are transformed back to the positive real line.}
\begin{tabular}{lcc}
& Out of Scanner & In Scanner \\ 
  \hline
 $b^{(a)}$ & 1.33 (.11) & 1.63 (.13) \\ 
 $b^{(n)}$ & 1.39 (.11) & 1.80 (.14) \\ 
 $b^{(s)}$ & 1.05 (.10) & 1.25 (.11) \\ 
 $A$ & 0.73 (.06) & 0.92 (.10) \\ 
 $v^{(e)}$ & 1.50 (.27) & 1.69 (.19) \\ 
 $v^{(c)}$ & 3.12 (.26) & 3.24 (.12) \\ 
 $\tau$ & 0.19 (.02) & 0.18 (.02) \\ 
\end{tabular}
\label{tab:forstmanmu}
\end{table}

\begin{figure}
\centering\includegraphics[scale=1.25]{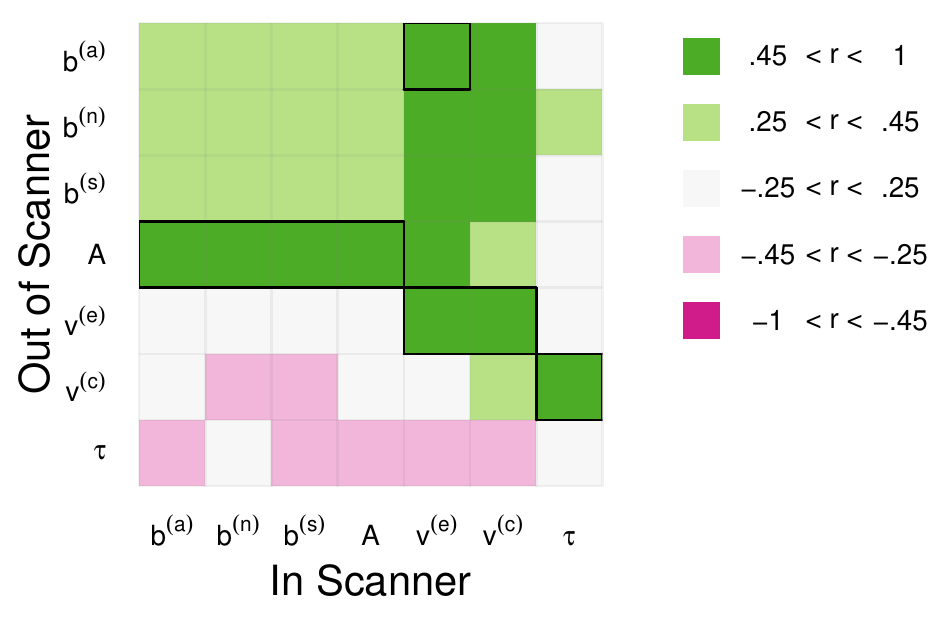}
\caption{Posterior means for the correlation matrix between parameters estimated for the out-of-scanner and in-scanner sessions of Forstmann et al.'s (2008) experiment. Correlations near zero are shown as white squares. Positive and negative correlations are shown by green and red shades, respectively. Cells enclosed by black borders are strongly reliable correlations, as indicated by having a posterior mean $\pm3$ or more standard deviations away from zero.}
\label{fig:forstmanncorr}
\end{figure}

Our main focus, however, is on the correlations between the parameters estimated from data recorded outside vs. inside the MRI scanner. The estimation method generates samples from the posterior distribution over the full covariance matrix. Appendix B shows the mean of these samples after transforming from the covariance matrix to the correlation matrix. Figure~\ref{fig:forstmanncorr} summarises just the most relevant section of the correlation matrix from Appendix B; it shows only the sub-section of the matrix with between-session correlations, the correlations of parameters estimated from out-of-scanner data with parameters estimated from in-scanner data. The figure summarises these correlations as a heatmap in which positive and negative correlations are represented by green and red colours respectively. Darker shades indicate stronger correlations, and cells enclosed by black borders have strong statistical reliability. 

The correlations between ``like'' parameters from different sessions are mostly as hypothesised, and easy to interpret. For example, all of the threshold-related parameters ($b^{(a)}$, $b^{(n)}$, $b^{(s)}$, and $A$) are positively correlated with each other between sessions, indicating that participants who made cautious decisions out of the scanner (high thresholds) also tended to make cautious decisions inside the scanner, and vice versa. The average magnitude of the correlations for threshold parameters ($r=.33$) is very similar to that reported by \cite{mueller2019individual} ($r=.39$).

The drift rate parameters ($v^{(e)}$ and $v^{(c)}$) are quite strongly correlated between sessions, with the exception of the error drift rate in-scanner paired with correct drift rate out of scanner. The average correlation between drift rates between sessions ($r=.42$) was almost double that reported by \cite{mueller2019individual}, which makes sense given that Forstmann et al.'s experiment was identical between sessions -- only the context changed. Non-decision time ($\tau$) was uncorrelated between sessions. 

The other correlations summarised in Figure~\ref{fig:forstmanncorr} are between ``unlike'' parameters, such as drift rates estimated out of the scanner correlated with thresholds measured in the scanner. These correlations are sometimes difficult to interpret. For example, the non-decision time ($\tau$) and correct accumulator drift rate ($v^{(c)}$) parameters estimated outside of the scanner correlate \emph{negatively} with almost all the other parameters estimated inside the scanner. This implies that people who were fast at the non-decision components of responding out of the scanner also tended to have high caution and large drift rates, when in the scanner. Others of the ``unlike'' correlations are easier to interpret. For example, participants who made cautious decisions outside the scanner (high $b^{(a)}$, $b^{(n)}$, $b^{(s)}$, and $A$) tended to perform the task well when inside the scanner (high $v^{(e)}$ and $v^{(c)}$).

\begin{figure}
\centering\includegraphics[scale=.8]{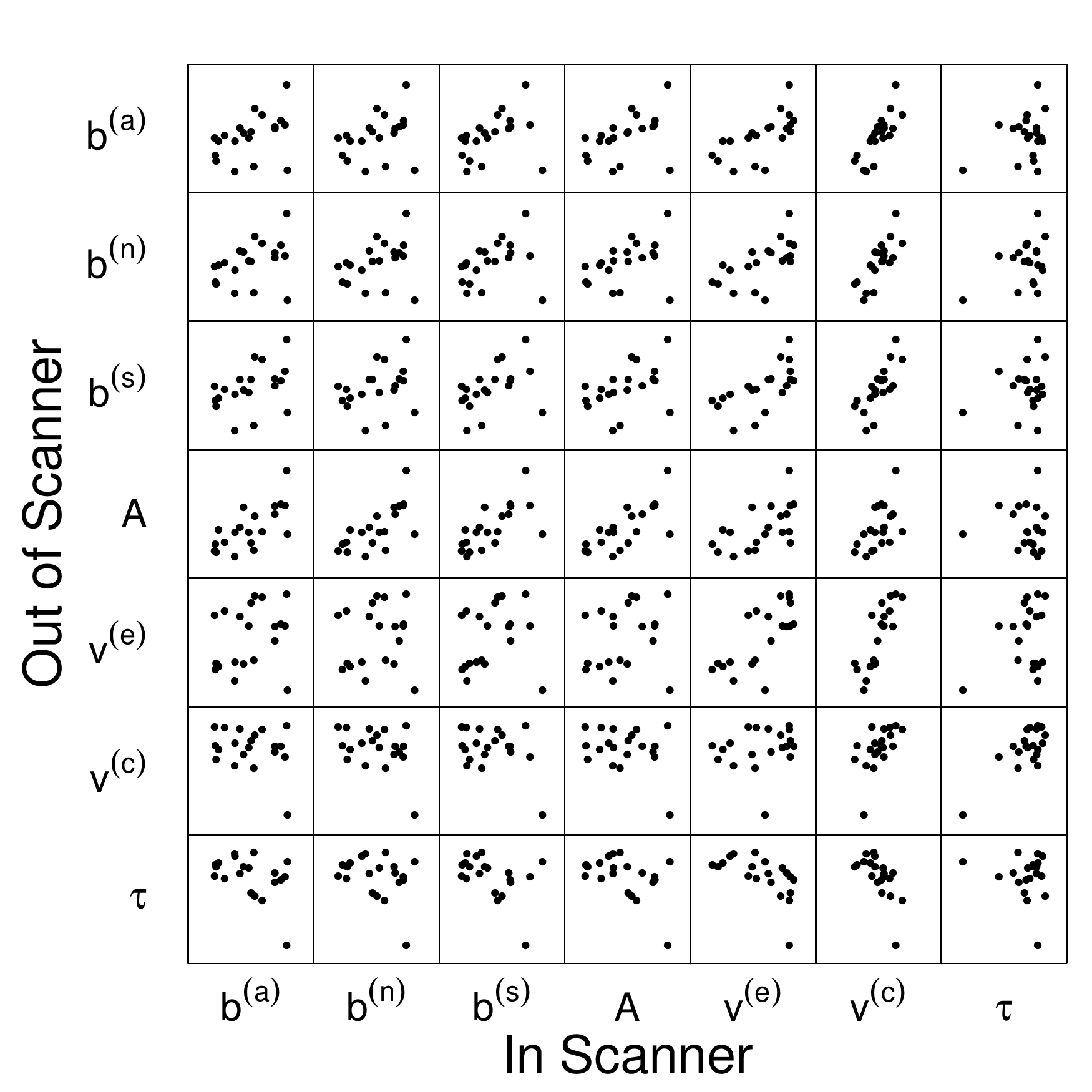}
\caption{Scatter plots of posterior mean estimates for the random effects parameters inside vs. outside of scanner.}
\label{fig:raneffforstmann}
\end{figure}

Only $n=19$ people participated in the experiment reported by \cite{forstmann2008striatum}, and it can be difficult to estimate correlation parameters with relatively small sample sizes -- despite the quite large number of data collected per person. The implication is clearly visible in Figure~\ref{fig:forstmanncorr}, where there are several cells with strong mean correlation (dark colors) that are still not strongly statistically reliable (no bounding boxes, indicating that the mean posterior correlation was less than 3 standard deviations from zero). Figure~\ref{fig:raneffforstmann} shows scatter plots corresponding to the correlations from Figure~\ref{fig:forstmanncorr}. Each panel in Figure~\ref{fig:raneffforstmann} has a symbol for each person in the experiment. Each symbol plots a point estimate for an in-scanner parameter vs. a point estimate for an out-of-scanner parameter. The point estimates are the means of the posterior distributions. Figure~\ref{fig:raneffforstmann} reveals that the relatively small number of participants contributed to the unstable correlations. For example, the negative correlations previously discussed, for out-of-scanner $\tau$ and $v^{(c)}$ with almost all in-scanner parameters, appear to be caused by an outlier (lowest value in each panel of the bottom two rows of Figure~\ref{fig:raneffforstmann}). The next experiment alleviates this difficulty by analysing a much larger sample of participants.

\subsubsection{Implications for model-based cognitive neuroscience}

Beyond this application, our method has the potential to enhance the reliability of model-based cognitive neuroscience research. A shortcoming of the field is that relatively few data can be collected while participants are inside a scanner, or while other neurophysiological recordings are taken. Given two testing sessions, one inside the scanner and another outside of the scanner, our method can improve the precision of the parameter estimates in both sessions, due to the borrowing of strength between and within tasks. 

We consider this a generalisation of the so-called joint-modelling framework that simultaneously estimates the parameters of a cognitive model (such as the LBA) and a neural model \citep[typically a GLM;][]{turner2013bayesian}. Joint modelling allows parameters estimated from one source (say, behavioural data) to influence parameters estimated from the other source (the neural data). Our approach tackles a trickier statistical problem, estimating the correlation between vectors of latent variables (parameters of cognitive models in different tasks, sessions, etc.) whereas to date joint modelling has been used to estimate the covariation between a set of latent variables (cognitive model parameters in one task) with a vector of data-transformed variables (beta-values in a GLM of the neural data). In this sense, our method is a generalisation of the joint modelling framework. It provides an avenue to estimate parameters of cognitive models from two behavioural sessions. This reduces uncertainty in the parameter estimates from the in-scanner session, in which there were fewer data, and also jointly models the in-scanner session and neural recordings, which can improve the estimation precision for the across-task covariance parameters.

\begin{figure}
\centering\includegraphics[scale=0.8]{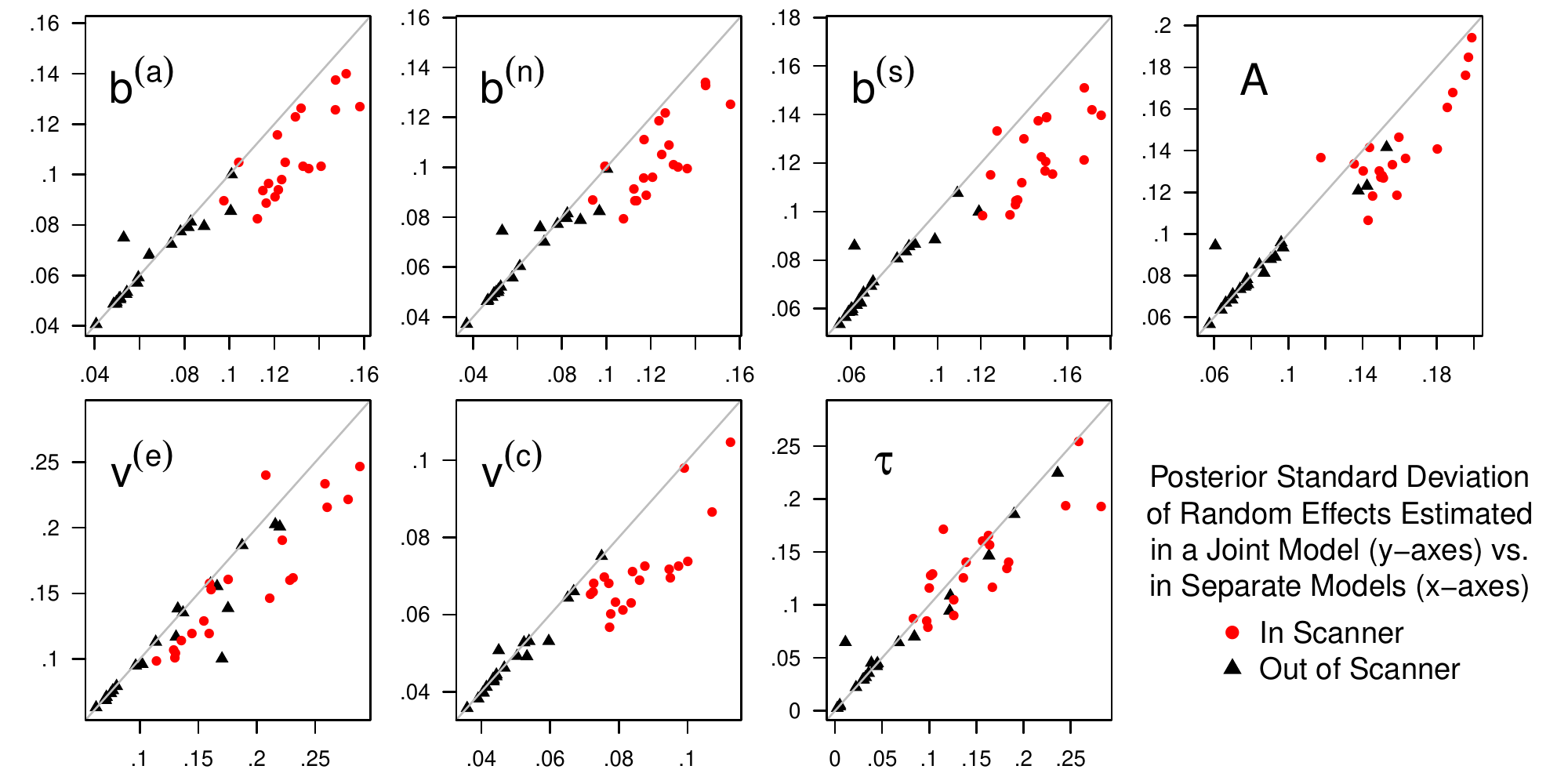}
\caption{Random effects are more precisely estimated in the joint model. Each panel represents one model parameter, and illustrates the precision with which the individual-subject random effects are estimated. Points show the posterior standard deviation for the jointly-estimated model ($y$-axes) vs. the independently-estimated models ($x$-axes). For data collected in the scanner (red circles), the posterior standard deviation is substantially smaller in the joint fit than in the independent fits. This improvement in precision is less apparent for the data collected out of the scanner (black triangles).}
\label{fig:precision}
\end{figure}

Figure~\ref{fig:precision} illustrates the improved estimation precision that can be gained by jointly modelling data from in- and out-of-scanner sessions. For each participant, we calculated the standard deviation of the samples drawn from the posterior distributions over their random effects -- both in and out of the scanner. Larger standard deviations correspond to poorer estimation precision. We then ran two new model analyses for comparison. These new analyses estimated the LBA model in the standard way: independently from the in-scanner and out-of-scanner data, maintaining the assumption of within-task correlations between parameters. We calculated the same standard deviation measures for the precision of the random effects estimated in these independent analyses. Each panel in Figure~\ref{fig:precision} shows the relationship between the precision of the jointly-estimated random effects ($y$-axes) and the precision of the random effects estimated in independent fits ($x$-axes). The comparison reveals three important outcomes. Firstly, the estimates were more precise -- with lower posterior standard deviations -- for the out-of-scanner data (black triangles) than the in-scanner data (red circles). This is expected given that participants contributed more than three times as much data out of the scanner than in the scanner. Secondly, estimation precision was better in the joint model than in the independent models (nearly 90\% of the symbols fall below the diagonal lines). Thirdly, the improvement in estimation precision was much more pronounced for the smaller data set (in-scanner) than the larger data set (out-of-scanner). For the in-scanner data, in red, the median change to the posterior standard deviation was 17\%. For the out-of-scanner data, the median improvement was just 1.2\%. This illustrates the point made above, that the benefits of modelling the covariance structure between tasks are most pronounced when there are relatively few data in some tasks. 

\subsection{Application 2: Correlations in Latent Processes Across Different Cognitive Tasks}

In the second experiment, participants completed three decision-making tasks in a single session. Compared with the experiment by \cite{forstmann2008striatum}, this experiment kept a constant context and environment for the participants, while the nature of the task varied. We also gathered data from many more participants ($n=110$). The differences between the three tasks means that lower correlations might be expected for parameters that are strongly dependent on the task; particularly drift rates. 

The tasks were a visual search task, a stop-signal task, and a match-to-memory task, which we abbreviate as ``search'', ``stop'', and ``match''. For the match task, we manipulated difficulty by changing the number of stimuli per trial (set sizes of 1, 2, or 3 objects). This manipulation was intended to change the speed and accuracy of decision-making, and to alter drift rates in the LBA model. The search and stop tasks had participants find a target stimulus, defined by a conjunction of colour and shape features, and then report the location of a small visual feature from the target. We manipulated the difficulty in the search and stop tasks by changing the properties of the distractor items. On some trials the target stimulus included a feature which was not present in any distractor stimulus; e.g., the target may have been red, while all distractors were green. These ``feature'' trials were the easiest for participants, and, by definition, all trials with just one distractor item were of this sort. For the trials with 3 or 7 distractor items, some were ``feature'' trials, but others were more difficult. The difficult trials are the ones in which both the features of the target were present in the distractors; e.g., searching for a red square amongst distractors that include a red circle and a green square. 

\begin{figure}
\centering\includegraphics[scale=1.1]{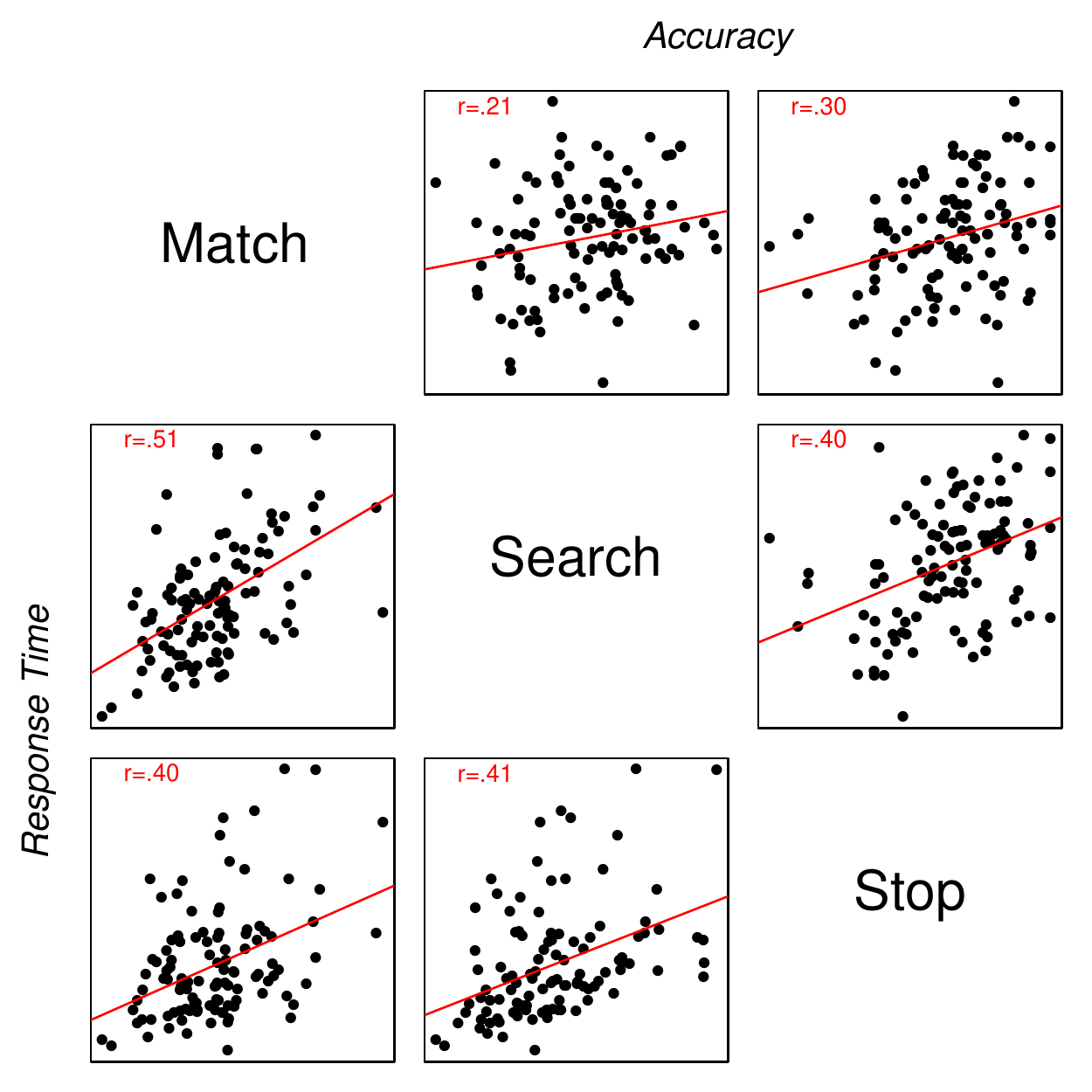}
\caption{Scatter plots of mean response time (RT; lower triangle) and accuracy (upper triangle) showing associations between performance in the three different tasks of the experiment. Accuracy is probit transformed. Red lines are regressions corresponding to the Pearson correlation coefficients shown in each panel.}
\label{fig:laura-scatter}
\end{figure}

Figure~\ref{fig:laura-scatter} demonstrates that there was some association in the observed performance across tasks. In the figure, each dot represents one participant's mean response time (RT; lower triangle panels) or mean accuracy (upper triangle panels). These means are plotted for one task (search, stop, or match) vs. another. For example, the lower-left panel plots mean RT in the match task on the $x$-axis against mean RT in the stop task on the $y$-axis. The correlations between tasks in RT were between $r=.40$ and $r=.51$, and for accuracy between $r=.21$ and $r=.40$. These correlations provide evidence that there is some commonality in performance between tasks which the cognitive modelling can strive to uncover and explain. 

Since the three tasks are different, the specification of the LBA model is not identical across them. This is different from the first application, to data from \cite{forstmann2008striatum}, in which the model was identical for the in-scanner and out-of-scanner sessions. For each of the three tasks, we constrained the model to use a single value for non-decision time ($\tau$) across conditions, and likewise a single value for the start-point variability ($A$) across conditions. The effect of display size was different in the three tasks. For the match task, blocks with larger display sizes were more difficult for participants than blocks with smaller display sizes. Reflecting this, we allowed different drift rates and different thresholds for the three different display sizes in the match task: $\left \{b^{(1)}, b^{(2)}, b^{(3)}\right \}$ for thresholds and $\left \{ v^{(1)}, v^{(2)}, v^{(3)} \right \}$ for drift rates. In the search and stop tasks, the effect of display size was modulated by the ``popout'' effect of feature (vs. conjunction) trials. We treated the feature trials as identical in the model, no matter which display size they used. Since all of the trials in display size 2 were feature trials, this implied thresholds of $\left \{b^{(f)}, b^{(4)}, b^{(8)}\right \}$, and drift rates $\left \{ v^{(f)}, v^{(4)}, v^{(8)} \right \} $. In most applications of evidence accumulation models, response thresholds are typically not allowed to vary with stimulus manipulations, such as display size. This is because it is implausible to imagine that decision-makers can adjust a response threshold contingent on some stimulus property, prior to making their decision about that stimulus. However, our experimental procedure provided participants with sufficient advance notice of the display size that thresholds could be plausibly adjusted. Finally, for the drift rates, we constrained the model to have just one parameter across all conditions to set the mean drift rate of the accumulator corresponding to the incorrect response, $v^{(e)}$. These model assumptions were the product of testing several other models, which were either simpler or more complex, and which either failed to capture important effects in the data or did not fit sufficiently better to justify the extra complexity.  

The model assumptions result in 9 unknown parameters for each participant, for each of the three different tasks. These parameters were estimated simultaneously across all three tasks. The vector of 27 log-transformed random effects was constrained to follow a multivariate normal distribution at the group level. Uninformed priors were assumed for the mean and covariance matrix of the multivariate normal, using the same settings as in the application to \citeowner{forstmann2008striatum} data.

Table~\ref{tab:triplemu} shows the estimated group-level parameters. Each entry gives the mean (with standard deviation in parentheses) for the posterior distribution over a group-level parameter, for one of the three tasks. For all three tasks, participants made more cautious decisions as display size increased; i.e., the estimated thresholds increased with display size, $b^{(1)} < b^{(2)} < b^{(3)}$ in the match task, and $b^{(f)} < b^{(4)} < b^{(8)}$ in the search and stop tasks. Decisions also became more difficult for the participants as display size increased in the search and stop tasks ($v^{(f)} > v^{(4)} > v^{(8)}$), although the corresponding effect was less clear in the match task.

\begin{table}[ht]
\centering
\caption{Mean (and SD) of the estimated marginal posterior distributions for the LBA mean parameters from the three tasks in the experiment; see text for details. For ease of interpretation, these parameters are transformed back to the positive real line.}
\begin{tabular}{lc|lcc}
 & Match &  & Search & Stop \\ 
  \hline
$b^{(1)}$ & 2.15 (.069) & $b^{(f)}$ & 1.71 (.049) & 2.74 (.171) \\ 
  $b^{(2)}$ & 2.34 (.073) & $b^{(4)}$ & 1.79 (.050) & 2.87 (.177) \\ 
  $b^{(3)}$ & 2.42 (.075) & $b^{(8)}$ & 1.95 (.055) & 3.04 (.180) \\ 
  A & 1.34 (.057) & A & 0.89 (.021) & 1.78 (.150) \\ 
  $v^{(e)}$ & 0.86 (.043) & $v^{(e)}$ & 1.19 (.056) & 0.77 (.084) \\ 
  $v^{(1)}$ & 2.94 (.049) & $v^{(f)}$ & 3.82 (.048) & 4.04 (.086) \\ 
  $v^{(2)}$ & 3.19 (.050) & $v^{(4)}$ & 3.62 (.046) & 3.96 (.094) \\ 
  $v^{(3)}$ & 2.79 (.044) & $v^{(8)}$ & 3.42 (.053) & 3.80 (.094) \\ 
  $\tau$ & 0.17 (.008) & $\tau$ & 0.22 (.008) & 0.23 (.008) \
\end{tabular}
\label{tab:triplemu}
\end{table}

\begin{figure}
\centering\includegraphics[scale=1.25]{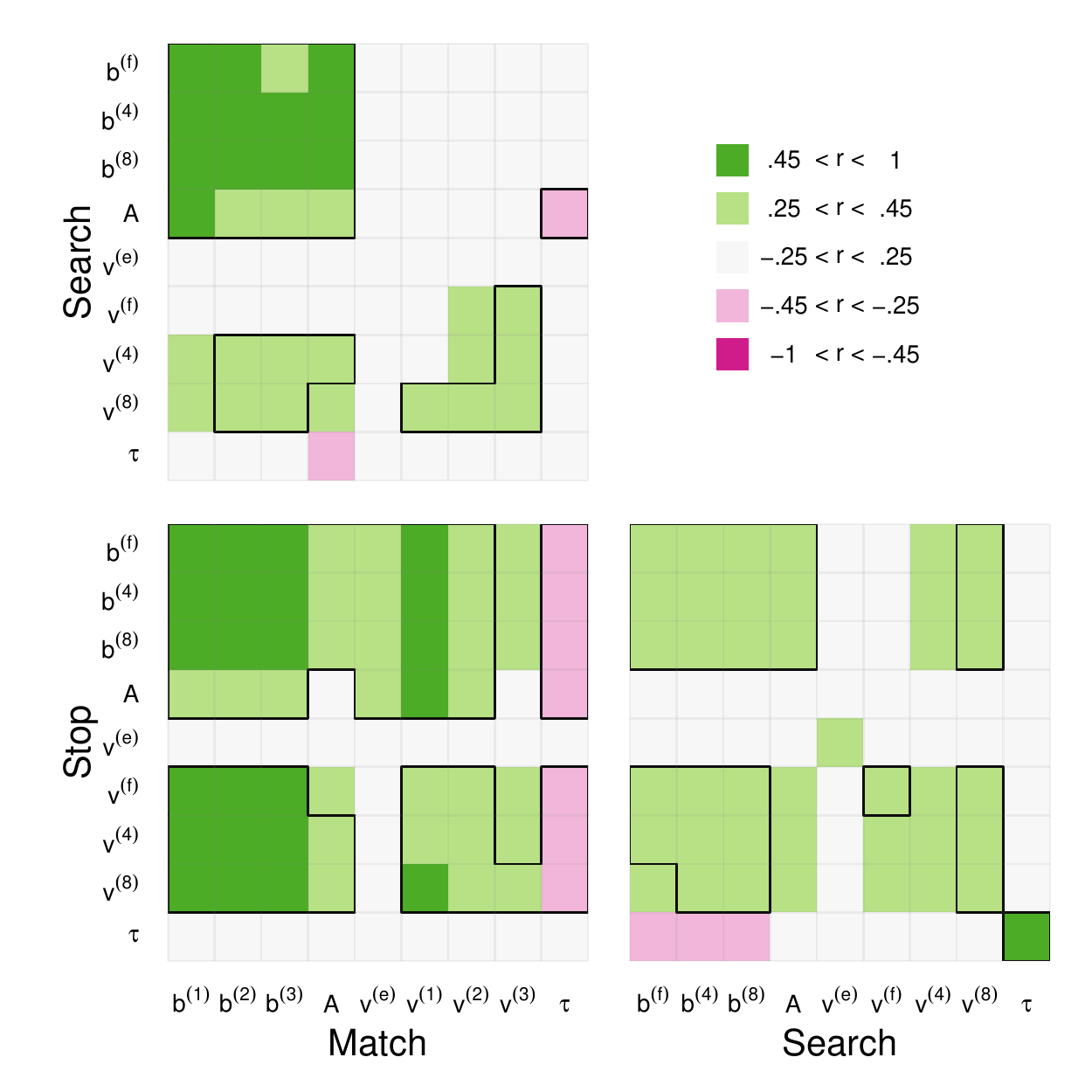}
\caption{Posterior mean estimates for the correlation matrix between parameters estimated for the three tasks (Match, Search, and Stop) in the experiment. Correlations near zero are shown as white squares. Positive and negative correlations are shown by green and red shades, respectively. Cells enclosed by black borders are strongly reliable correlations, as indicated by having a posterior mean $\pm3$ or more standard deviations away from zero.}
\label{fig:heatmaptriple}
\end{figure}

Figure~\ref{fig:heatmaptriple} uses the same plotting format as used in Figure~\ref{fig:forstmanncorr}, so that red and green shades indicate negative and positive parameter correlations, respectively, with darker shades corresponding to stronger correlations. The positioning of the panels is the same in Figure~\ref{fig:heatmaptriple} as for the lower triangle of Figure~\ref{fig:laura-scatter}. Appendix B gives the correlation values corresponding to Figure~\ref{fig:heatmaptriple}.

The dark green patches on the left-hand sides of the two left panels indicate that the threshold estimates for the match task correlate positively with threshold estimates from the other two tasks, and also with correct-accumulator drift rates for the stop task. The light-shaded horizontal and vertical sections for parameter $v^{(e)}$ suggest that the drift rates for the incorrect accumulator have low or no correlations with any other estimates. This result is consistent with the idea that error drift rates are noisy to estimate, especially when accuracy is high. The non-decision time parameter ($\tau$) from the search task does not correlate strongly with any other parameters except for the non-decision time parameter for the stop task. The two non-decision time parameters for those two tasks correlated strongly (bottom right element in the lower right panel), which makes sense given that the stop task and search task used identical response rules -- participants responded to the side of the target stimulus which showed a small gap. The non-decision time parameter for the match task does not correlate with those from the other tasks, which also makes sense because the match task required a different response rule; match to memory, which presumably requires different encoding than the gap identification, and also a different mapping to the response key.

\subsubsection{Implications for test batteries}

The second application demonstrates that our approach can identify relationships between the latent cognitive processes involved in different tasks. In this application, the tasks involved finding a target among distractors, decisions in the context of response inhibition, and matching stimuli to previously-remembered referents. Given we had a considerable number of decisions per task, it may have been possible -- and simpler -- instead to independently estimate the parameters of the cognitive model for each task, and then conduct pairwise correlations between the parameter estimates. Even in this many-trials context, we believe our method has important uses. For example, it provides a new method for assessing test-retest reliability of model parameters across testing occasions.

Nevertheless, in many contexts it is impossible to independently estimate cognitive models for each task. For example, in clinical samples it is common for participants to complete many different tasks -- up to 10 in a session -- with very few trials per task. Performance in such ``test batteries'' including the BACS \citep{kaneda2007brief}, CANTAB \citep{robbins1996neural} and MiniMental \citep{folstein1975mini} are used to inform important clinical decisions about cognitive functioning in patients, and are often used in research to assess whether an intervention is effective at improving cognition \citep{john2017successful,demant2015effects}. It is therefore of practical and theoretical importance that the inferences drawn from test batteries are based on precise measurement. However, these inferences are typically based on composite scores derived from summary statistics such as the mean RT or number of lapses, calculated from small data samples. There are likely to be substantial within-subject correlations across the multiple tasks, though current treatments ignore those, and treat the tests independently. Our method allows us to explicitly model the dependence across tasks, which provides more precise parameter estimates, and the benefits of more psychologically sensible assumptions about shrinkage \cite[see][]{rouder2019psychometrics}. Explicitly modelling the correlations between tasks also opens up theoretically interesting possibilities, such as testing cognitive models of performance as elements of larger test batteries. This has been inaccessible to cognitive modelling, at least in applied domains, owing to the issue of few data per task. There are likely to be important issues that need to be resolved in future, in order to make that work. \cite{rouder2019most} discuss how methodological differences between cognitive tasks and psychometric tests emphasise different psychometric properties which can make it difficult to draw consistent inferences between them \cite[but see also][]{kvam2019testing}.

\subsection{Simulation Study}

The two applications identified statistically reliable covariances between the individual-subject parameters, i.e. random effects, across different tasks or different sessions. These relationships are important for methodological reasons, but also scientifically, in that they reveal stable trait-level properties of people. We conducted a simulation study to increase confidence in such scientific conclusions. The goals of the simulation study were to establish that, given good input data, the covariance-modelling method we have developed: (a) accurately recovers a known covariance structure in simulated data; (b) does not support misleading inference about reliably non-zero covariance in data simulated with zero covariance; and (c) reliably supports inference of non-zero covariance in data simulated with non-zero covariance. 

We ran three versions of the simulation study. The studies simulated data from an experiment based on that of \cite{forstmann2008striatum}, but with $S=100$ participants each contributing $n=1,000$ trials in each of the in-scanner and out-of-scanner sessions. The three versions of the simulation study varied only in the covariance parameters used to generate the data. For all three versions, the population mean parameters and the associated variance parameters used to generate data were matched to the mean values estimated from the fits to Forstmann et al.'s data; see Table~\ref{tab:forstmanmu}. 
For the first and second versions, the covariance parameters for within-session random effects were also matched to the mean values estimated from data; see Table~\ref{tab:forstmann-cor-appendix}. For example, the data-generating parameter for the covariance between $b^{(a)}$ from the in-scanner session and $\tau$ from the in-scanner session was set to the mean of the posterior samples for that parameter, from Application 1. The two versions differed in how they set the data-generating parameters for the covariance between in- and out-of-scanner parameters; there are 49 such parameters in each version. In the first version, these were also matched to the mean values estimated from real data. In the second version, all between-session covariance parameters were set to zero; i.e., random effects for in-scanner and out-of-scanner sessions were independent. For the third version, we set all within and between-session covariance parameters to non-zero values; specifically, covariance values which implied correlations of $r=.8$ between pairs of random effects, in the data-generating process.


\begin{figure}
\centering\includegraphics[scale=1]{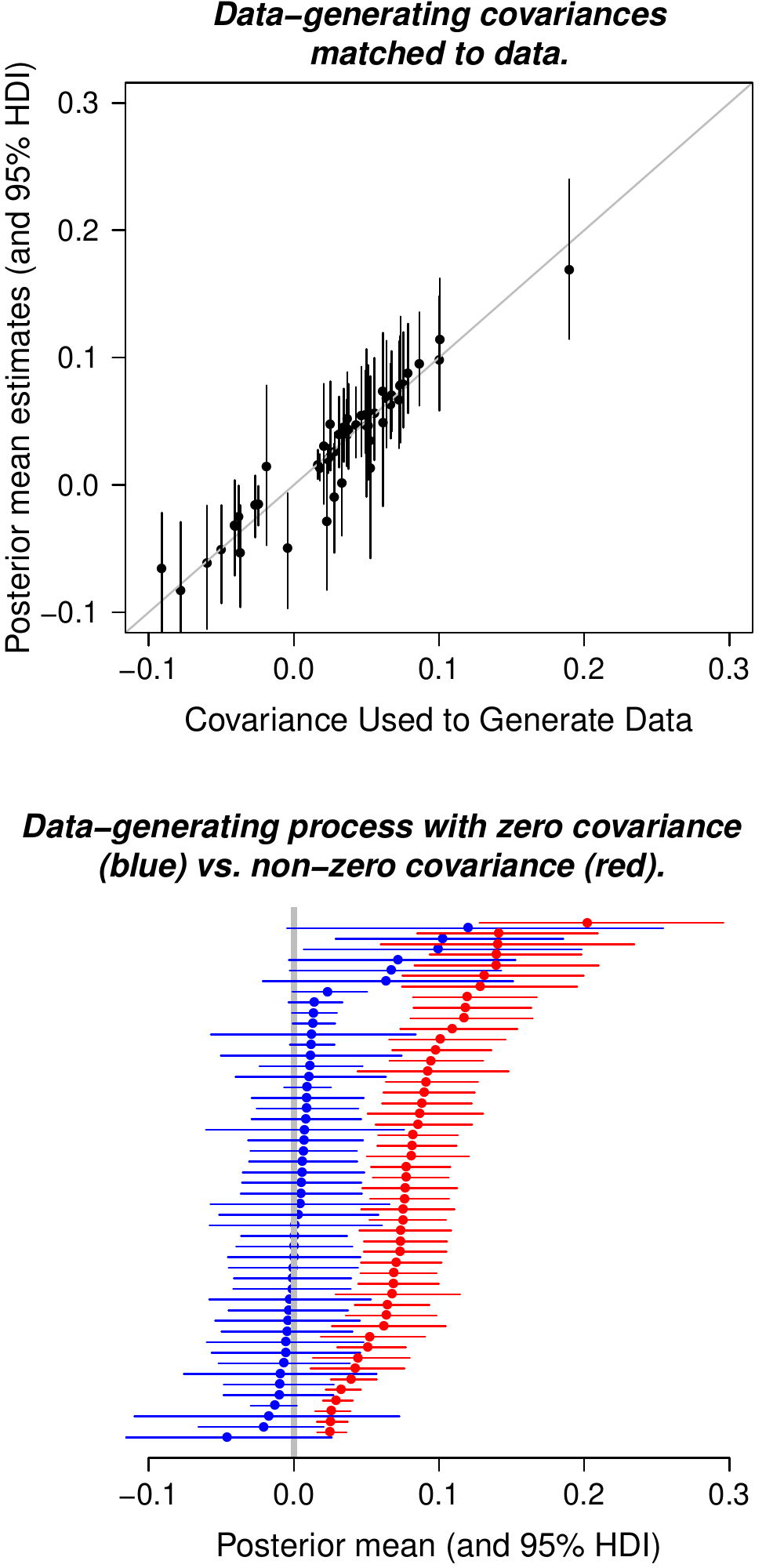}
\caption{Covariance recovery simulation. For the first version of the simulation study (top panel), the covariance values used to generate the data ($x$-axis) were set to the mean estimates from the data in Application 1. The 95\% posterior credible intervals estimated from the simulated data ($y$-axis) include the data-generating value in almost every case. In two follow-up versions of the simulation study (lower panel), the data-generating process assumed independent random effects for the in- and out-of-scanner data (blue -- zero covariance) or uniformly non-zero dependence (red). The estimated posterior distributions include zero for almost every element of the covariance matrix when the processes are independent (blue), and exclude zero for every case where the processes are dependent (red).}
\label{fig:simulationrecovery}
\end{figure}

The top panel of Figure~\ref{fig:simulationrecovery} illustrates the results from the first version of the simulation study. This panel shows the data-generating covariance parameters ($x$-axis) and the values recovered for these parameters ($y$-axis, with means and 95\% credible intervals). Matching the values estimated from real data, the covariance parameters used to generate the simulated data include some that are close to zero, and some that are quite large (corresponding to the correlations reported in Figure~\ref{fig:forstmanncorr}). The recovered posterior distributions include the data-generating values inside their credible interval in almost every case. This confirms the first aim of the simulation study.

The lower panel of Figure~\ref{fig:simulationrecovery} illustrates results from the second and third versions of the simulation study. The blue symbols and lines show posterior means and 95\% credible intervals for the covariance parameters estimated from the second version, in which the corresponding data-generating covariance parameters were all zero. In almost every case, the recovered posterior distributions include zero (the vertical gray line). This confirms the second aim of the simulation study, showing that the model reliably infers independent random effects when that is appropriate. The red symbols and lines show the posteriors estimated when the data-generating covariance parameters were all non-zero. In this case, all of the estimated credible intervals are above zero. This confirms the third aim of the simulation study, showing that the approach reliably detects correlated random effects between sessions, when that is appropriate.

\section{Conclusions}

Our article develops a statistically principled approach to estimate the degree of association between the latent cognitive processes that drive performance across tasks, contexts, and time. Most previous research assessing parameter correlations across testing occasions has been restricted to estimating the parameters of cognitive models independently for each test session, and then correlating the pointwise estimates of those parameters in a second-step analysis. Such an approach has conceptual and statistical shortcomings. 

Conceptually, existing approaches start with the assumption that cognitive processes are independent over tasks, contexts and time. This is surely not true, and is inconsistent with an assumption underlying all psychological research that there is some non-zero degree of stability in psychological processing across contexts and over time. It is this consistency we aim to uncover and use as a basis for generalisation. Our method allows us to identify the similarity in cognitive processing between different testing occasions, without making the (implicit) assumption that the latent drivers of observed performance are independent across testing occasions. 

Statistically, existing approaches are over-confident: they use point estimates of the parameters from independent model fits to each task. This assumes the parameters of participants are known with certainty within a task, which is never true when analysing data; providing the machinery to deal with this uncertainty is one of the primary advantages of Bayesian methods. Furthermore, with existing approaches there are just two ways to assess relatedness in parameters across testing occasions: assuming independence or equivalence; i.e., tying parameters across conditions or tasks. Where it is a priori unclear which parameters can be assumed to be constant across conditions or tasks, we can get stuck with independent fits, or even without being able to progress. Estimating a dependent pair of parameter vectors allows for a ``soft'' version of tying parameters across conditions. Parameters which are related will then show up as correlated, and statistical borrowing of strength will take place via the covariance matrix. New work reported by \cite{kvam2019testing} takes a related approach to ours, aiming to borrow information across different testing tasks in a clinical sample, they demonstrate improved estimation precision in their joint modelling approach.

The analyses of data from \cite{forstmann2008striatum} showed that estimating parameters jointly across correlated tasks (or sessions) can improve the precision of subject-level estimates. This can be important when there are limitations on the number of data which are available in some tasks, for example, due to limitations in the number of stimuli available or in the persistence of the participants. When the sample size is very different between the sub-tasks, the improvement in estimation precision gained by jointly modelling the tasks and their covariance will be greatest for the tasks with fewest data. Future work may explore ways of exploiting this for maximum benefit. For example, when one particular sub-task is of high value, but has strict limits on its sample size, estimation precision in that sub-task may be improved by collecting more data on other, related, tasks.

\newpage
\section{Open Practices Statement}

The two applications cover a previously published data set \citep{forstmann2008striatum} and a new experiment that was not preregistered. Data and code for both applications are available at \url{osf.io/rf8nd}.

\newpage
\bibliography{uoncoglabshort}

\newpage

\appendix

\section{The New Experiment}

\subsection{Method}
\subsubsection{Design}
The experiment used a 3 (task) $\times$ 3 (set size) within-subjects design: all three tasks had a three-level manipulation of the number of items in the stimulus array (set size). In the search task, participants were required to look for a target (always present) amongst one, three or seven distractors (implying search set sizes of 2, 4, and 8). The stop-signal task was identical to the search task except that on 25\% of trials a stop-signal was presented after the onset of the search array. The time between the onset of the search array and the stop-signal (called the stop-signal delay) was dynamically adjusted for each participant and each set size, using a staircase algorithm. In the match task participants were required to identify if the currently presented stimulus set was a match (the same shapes and colours) or not a match (at least one difference) to the stimulus set presented on the previous trial. The number of stimuli present on screen in each trial was either one, two, or three, and this was manipulated between blocks of trials. Response time and the response itself were recorded for all trials.

\subsubsection{Participants}
Participants were students from first and second year psychology courses at the University of Newcastle who received course credit for their participation. Informed consent was obtained for all participants. Participants had the opportunity to complete the task online ($N=106$) or in a lab ($N=81$).

Although 187 students participated in the study, only 148 participants are included in the combined analysis. Participants were excluded if they had greater than 0.05\% of non-responses due to ``too fast'' or ``too slow'' feedback cut offs, as defined in the procedure ($n=8$ match, $n=10$ search, $n=13$ stop) or had accuracy lower than 75\%, 85\% or 90\% for the match, search and stop-signal tasks respectively ($n=17$ match, $n=8$ search, $n=27$ stop). The exclusion criteria were set by investigating the data and removing outliers indicating the participant was performing considerably worse at the task than the bulk of the other participants. This resulted in $n=145$ complete data sets for the match task, $n=157$ for the search and $n=133$ for the stop-signal. There were $n=110$ participants who had valid data for all three tasks. These were the participants used in all analyses.

\subsubsection{Materials and Stimuli}
All three tasks were written in JavaScript and HTML5. Although it was impossible to keep screen size and resolution identical across subjects who completed the task online, the relative size and positioning of stimuli was constant. Instructions at the beginning of the experiment required participants to alter the zoom settings to ensure maximum consistency in the displays across participants. On a 1920x1080 resolution and 13.3 inch screen, with the participants 60cm away from the screen, each shape subtended approximately $1^\circ$ of visual angle, and was approximately $5^\circ$ of visual angle away from the center of the screen. The fixation point was a small cross subtending much less than $1^\circ$ of visual angle.
Stimuli only ever appeared in eight different locations, representing equally-spaced points around a circle $5^\circ$ in radius. For stimulus displays with fewer than eight stimuli, locations were sampled randomly without replacement from the eight possible positions. 

\begin{figure}[!ht]
\centering\includegraphics[width=1\linewidth]{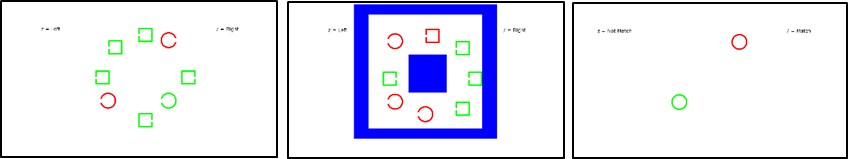}
\caption{Screenshots from: (left) the search task, set size eight, with the green circle as the target; (middle) the stop-signal task, set size eight, target red square, with the stop-signal present; and (right) the match task, set size two.}
\label{TaskScreenshots}
\end{figure}

The search arrays for all three tasks used just four stimuli: a red circle, green circle, red square, and green square; see Figure \ref{TaskScreenshots}. A small gap in the shape, on either the left or right side, was used as the decision stimulus in the search and stop-signal tasks. There was no gap in any stimulus during the match task. The three colours used (red and green for the stimuli, and blue for the stop-signal) were presented at the maximum intensity of their respective hue in the computer's RGB colour model.  

\subsubsection{Procedure}

Participants completed all three tasks in one sitting with opportunities for self-timed breaks within and between tasks. Participants were randomly allocated to one of the six possible task orders. Each task contained on-screen instructions with examples, followed by a series of practice trials and then three experimental blocks with a fixed number of trials each.

For the {\bf search task}, participants were first presented with instructions that identified which of the four stimuli would constitute their target stimulus; e.g., ``search for a red square''. The target was randomly allocated to participants at the start of the task and remained consistent for the duration of that task. This process also occurred in the stop-signal task and thus the target could change across tasks but not within a task. Participants were informed that all shapes have a gap in the left or the right side and that once participants had located the target they should indicate via the ``z'' and ``/'' keys if the gap was on the left or right side of the target respectively. Participants were told to respond as quickly as possible. 

There were three blocks of 200 trials each, with 10 practice trials at the start. At the beginning of each trial a fixation cross was presented for 700ms. This was then replaced by the search array. The location of the target and the target gap side were randomly chosen at the beginning of each trial. The number of distractors, their shape, colour, location and gap side were also randomly chosen at the start of each trial. A trial concluded after a response. If a response was faster than 250ms or slower than 2,000ms, then feedback of ``TOO FAST'' or ``TOO SLOW'' was provided, displayed for 1,500ms or 5,000ms, respectively. Participants also received accuracy feedback for the first 10 experimental trials. This feedback was presented for 1,000ms and 2,500ms for correct and error responses respectively.

{\bf Stop-signal tasks} typically have a very simple, almost automatic task for most trials in which participants rapidly press a key to respond on each trial, these are called the ``go'' trials. A stop-signal appears during the other trials (the ``stop'' trials), after some delay from the onset of the trial, and participants must withhold their response. In our stop-signal task, the go trials were identical to the search task. All details of the search task were identical except that in the instructions participants were shown a large blue square (see Figure \ref{TaskScreenshots}) and told ``when you see this symbol DO NOT RESPOND''. They were reminded to respond as quickly as possible when the symbol is not presented to ensure the easy, automatic response style.

Each trial had an independent and identical 25\% probability of being a stop-signal trial. At the beginning of the experiment the stop-signal delay (SSD; the time between the presentation of the stimuli and the presentation of the stop-signal) was set to zero across all set sizes, and then adjusted by a staircase procedure independently for each set size. After each correct inhibition, SSD was increased by 50ms (thus making it harder to inhibit) and after each failed inhibition, SSD was decreased by 50ms, with a minimum of zero. These staircases converge to the SSDs corresponding to 50\% successful inhibition in each set size. Figure \ref{TaskScreenshots} shows the stop-signal task, consisting of a blue square in the center of the screen, inside the eight pointed star of shapes (subtending approximately $5^\circ$ x $5^\circ$ of visual angle), and a larger outline of a square outside the eight pointed star (approximately $15^\circ$ x $15^\circ$ visual angle in size, width of outline approximately $1.75^\circ$ visual angle. To reduce so-called ``trigger failure'' \cite[see][]{matzke2017failures}, the stop-signal was maintained on screen until the end of the trial. To also prevent participants ignoring the stop-signal they were provided with feedback after every stop trial. Successful inhibitions produced ``Good stopping!'' while failed inhibitions resulted in ``You should have stopped''.  

\begin{figure}[!ht]
\centering\includegraphics[width=1\linewidth]{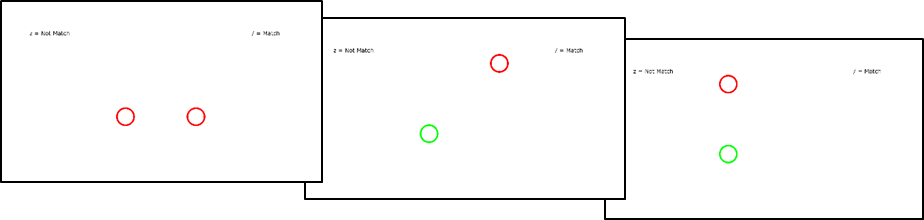}
\caption{Example trial sequence from set size two in the match task. The correct response for the second screen would be ``non-match'', and the correct response for the third screen would be ``match''. Both colours and shapes must be the same, but location does not matter.}
\label{MatchScreenshot}
\end{figure}

The {\bf match task} commenced with on-screen instructions that informed the participant green and red circles and squares would be shown on a trial and they needed to remember what they saw for the next trial. At first, they would only be shown one shape to remember, then two and then three shapes. If the stimulus array on any trial consisted of the same stimuli as the previous trial (i.e., with the same shapes and colours), then participants were to press a key indicating match (``/'' key). If any shape or colour differed, participants were to indicate a non-match (``z'' key, as seen in Figure \ref{MatchScreenshot}). Participants were explicitly instructed that the position of the stimuli on screen was irrelevant. 

Unlike the other two tasks, set size was not randomised from trial-to-trial for the match task, because match vs. non-match is not well defined for arrays of unequal size. Instead, set sizes occurred in blocks in a fixed order from one to three, with 100 trials per block. Half of the trials were randomly selected to be matching trials, and the other half were non-matching trials. During the experiment, if the upcoming trial was a match trial, the stimuli were kept fixed from the previous trial, however their locations were randomly resampled. If the upcoming trial was not a match then both the stimuli and location were randomly sampled, subject to the constraint that a match did not occur by chance. The feedback was the same as for the search task; however, the timeout for ``too long'' responses was 3,000ms and the accuracy feedback continued for the duration of the experiment in the match task. These changes in feedback were implemented after pilot testing, to allow for the greater difficulty of the match task.

\subsection{Results}

\begin{figure}[!ht]
\centering\includegraphics[width=1\linewidth]{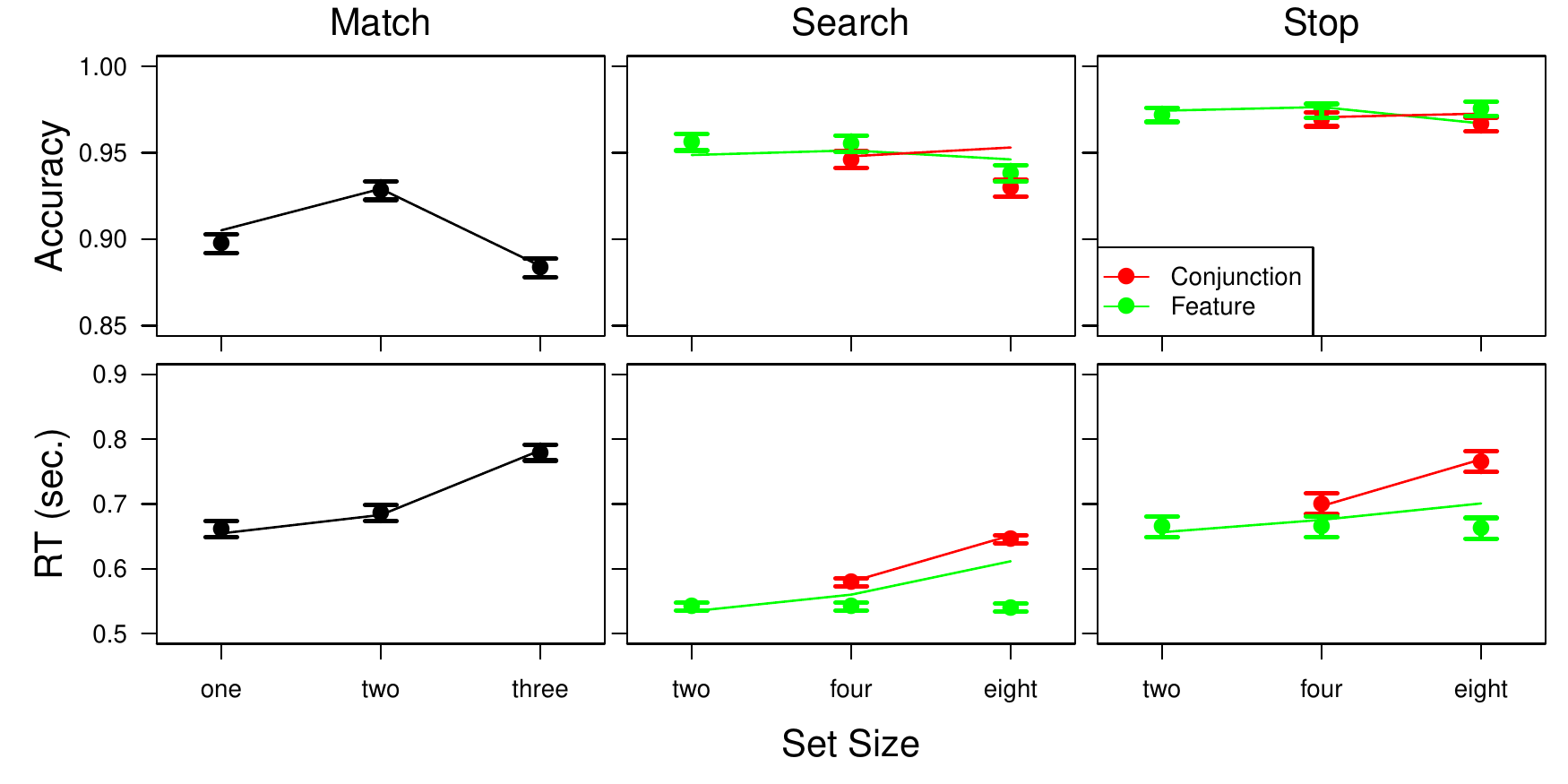}
\caption{Accuracy (top row) and response time (RT; bottom row) for the three tasks (columns) in the experiment. Accuracy and median RT were calculated for each participant and each condition. Lines show the averages of these across conditions for data, and symbols show the same but for posterior predictive data generated by the LBA model described in the main text. Red and green colours indicate trials in which the features of the target stimulus appeared in the distractor items (``conjunction'') and when they did not (``feature''), respectively. Error bars show $\pm1$ standard error of the mean for differences between participants in the model fits.}
\label{fig:descriptivestriple}
\end{figure}

Figure~\ref{fig:descriptivestriple} shows the mean response time (RT) and accuracy for different conditions, and for each of the three tasks. 

Bayesian repeated measures ANOVAs \citep{morey2013bayesfactor} were conducted on the mean RT and accuracy data for each set size, separately for the three tasks. Figure \ref{fig:descriptivestriple} shows that there is a strong effect of set size on RT for all three tasks, reflected in the RT Bayes factors in Table \ref{RTAccTable}. This is not true of accuracy. Although the match task shows an effect of set size on accuracy, for both the search task and the stop task accuracy did not change reliably across set size; see Table \ref{RTAccTable}. This may be due to ceiling effects, as in both those tasks average accuracy is around 95\% or above for all conditions. Even in the match task, accuracy does not decrease monotonically over set size as expected. Instead, there is a small increase from the smallest set size (one) to the medium set size (two) and then decreases to the largest set size (three). This is most likely due to practice effects, as the conditions were administered in blocks with increasing set size.

\begin{table}[h]
\centering
\caption{Bayes factors in favour of a model including an effect of set size vs. a null model, for the three tasks, for mean RT and accuracy.}
\begin{tabular}{l l l}
Task & Variable & Set Size BF$_{10}$ \\
\hline
 Search & RT & $>10^{6}$\\
  & Accuracy & 5.8  \\ \hline
 Stop & RT & $>10^{6}$\\
  & SSRT & $>10^{6}$\\
  & Accuracy & 0.060 $*$ \\ \hline
Match & RT &  $>10^{6}$\\
  & Accuracy & $>10^{5}$\\
\end{tabular}
\label{RTAccTable} 
\end{table}

The RT data are simpler to interpret. The only noteworthy point is that RT is substantially faster for the search task than both the stop and match tasks. As the RTs presented in the table and figure for the stop task are the RTs from the go trials (which are identical to trials in the search task), this increase in RT from the search to the stop task suggests that participants slow their responses when a stop is introduced to the task.

\newpage 

\section{Estimated Correlation Matrices }

Tables \ref{tab:forstmann-cor-appendix} and \ref{tab:triplecor} show the full correlation matrices for the two applications of the new method reported above.

\begin{table}[ht]
\centering
\caption{The lower triangle of the estimated posterior mean of the correlation matrix for Forstmann et al.'s (2008) experiment; the main text gives further details. The lower-left square shows between-session correlations; these are also shown in the main text as a heat map. The triangles on the diagonal blocks show within-session correlations.}
\begin{tabular}{ll|lllllll|llllll}
  & & \multicolumn{7}{c}{Out of Scanner} & \multicolumn{6}{|c}{In Scanner} \\
 & & $b^{(a)}$ & $b^{(n)}$ & $b^{(s)}$ & A & $v^{(e)}$ & $v^{(c)}$ & $\tau$ & $b^{(a)}$ & $b^{(n)}$ & $b^{(s)}$ & A & $v^{(e)}$ & $v^{(c)}$ \\ 
  \hline
  \multirow{6}{*}{\rotatebox[origin=c]{90}{Out of Scanner}} & $b^{(n)}$ & .97 &  &  &  &  &  &  &  &  &  &  &  &  \\ 
  & $b^{(s)}$ & .94 & .9 &  &  &  &  &  &  &  &  &  &  &  \\ 
  & A & .75 & .74 & .67 &  &  &  &  &  &  &  &  &  &  \\ 
  & $v^{(e)}$ & .74 & .71 & .74 & .32 &  &  &  &  &  &  &  &  &  \\ 
  & $v^{(c)}$ & .64 & .66 & .59 & .2 & .66 &  &  &  &  &  &  &  &  \\ 
  & $\tau$ & -.67 & -.65 & -.66 & -.48 & -.51 & -.41 &  &  &  &  &  &  &  \\ 
  \hline
  \multirow{7}{*}{\rotatebox[origin=c]{90}{In Scanner}} & $b^{(a)}$ & .38 & .33 & .36 & .59 & .2 & -.22 & -.32 &  &  &  &  &  &  \\ 
  & $b^{(n)}$ & .29 & .28 & .26 & .59 & .08 & -.31 & -.23 & .92 &  &  &  &  &  \\ 
  & $b^{(s)}$ & .35 & .29 & .38 & .57 & .17 & -.28 & -.32 & .92 & .87 &  &  &  &  \\ 
  & A & .39 & .35 & .37 & .62 & .17 & -.22 & -.34 & .9 & .87 & .87 &  &  &  \\ 
  & $v^{(e)}$ & .56 & .53 & .54 & .47 & .57 & .15 & -.41 & .73 & .65 & .66 & .63 &  &  \\ 
  & $v^{(c)}$ & .59 & .57 & .54 & .39 & .55 & .41 & -.43 & .46 & .38 & .38 & .4 & .63 &  \\ 
  & $\tau$ & .24 & .29 & .16 & -.03 & .21 & .61 & -.12 & -.49 & -.48 & -.55 & -.39 & -.26 & 0 \\ 
  \label{tab:forstmann-cor-appendix}
\end{tabular}
\end{table}


\begin{landscape}

\begin{table}[ht]
\centering
\caption{The lower triangle of the estimated posterior mean of the correlation matrix for our experiment; the main text gives further details. The three lower-left squares show between-task correlations; these are also shown in the main text as heat maps. The three triangles on the diagonal blocks show within-task correlations.}
\resizebox{22cm}{!}{\begin{tabular}{cl|ccccccccc|ccccccccc|cccccccc}
&   & \multicolumn{9}{c}{Match Task} & \multicolumn{9}{|c|}{Search Task} & \multicolumn{8}{|c}{Stop Task} \\
  & & $b^{(1)}$ & $b^{(2)}$ & $b^{(3)}$ & A & $v^{(e)}$ & $v^{(1)}$ & $v^{(2)}$ & $v^{(3)}$ & $\tau$ & $b^{(f)}$ & $b^{(4)}$ & $b^{(8)}$ & A & $v^{(e)}$ & $v^{(f)}$ & $v^{(4)}$ & $v^{(8)}$ & $\tau$ & $b^{(f)}$ & $b^{(4)}$ & $b^{(8)}$ & A & $v^{(e)}$ & $v^{(f)}$ & $v^{(4)}$ & $v^{(8)}$ \\ 
  \hline
  \multirow{8}{*}{\rotatebox[origin=c]{90}{Match Task}} & $b^{(2)}$ & .95 &  &  &  &  &  &  &  &  &  &  &  &  &  &  &  &  &  &  &  &  &  &  &  &  &  \\ 
&  $b^{(3)}$ & .9 & .93 &  &  &  &  &  &  &  &  &  &  &  &  &  &  &  &  &  &  &  &  &  &  &  &  \\ 
&  A & .89 & .87 & .85 &  &  &  &  &  &  &  &  &  &  &  &  &  &  &  &  &  &  &  &  &  &  &  \\ 
&  $v^{(e)}$ & -.03 & -.09 & -.2 & -.21 &  &  &  &  &  &  &  &  &  &  &  &  &  &  &  &  &  &  &  &  &  &  \\ 
&  $v^{(1)}$ & .16 & .21 & .19 & .06 & .14 &  &  &  &  &  &  &  &  &  &  &  &  &  &  &  &  &  &  &  &  &  \\ 
&  $v^{(2)}$ & .08 & .22 & .19 & .05 & -.08 & .57 &  &  &  &  &  &  &  &  &  &  &  &  &  &  &  &  &  &  &  &  \\ 
&  $v^{(3)}$ & .13 & .22 & .33 & .2 & -.19 & .27 & .5 &  &  &  &  &  &  &  &  &  &  &  &  &  &  &  &  &  &  &  \\ 
&  $\tau$ & -.39 & -.32 & -.23 & -.16 & -.32 & -.06 & .16 & .38 &  &  &  &  &  &  &  &  &  &  &  &  &  &  &  &  &  &  \\ 
  \hline
  \multirow{9}{*}{\rotatebox[origin=c]{90}{Search Task}} &  $b^{(f)}$ & .48 & .46 & .42 & .47 & .18 & .08 & -.01 & .09 & -.13 &  &  &  &  &  &  &  &  &  &  &  &  &  &  &  &  &  \\ 
&  $b^{(4)}$ & .53 & .51 & .47 & .52 & .17 & .1 & .01 & .09 & -.16 & .99 &  &  &  &  &  &  &  &  &  &  &  &  &  &  &  &  \\ 
&  $b^{(8)}$ & .52 & .51 & .47 & .5 & .16 & .15 & .07 & .09 & -.17 & .97 & .98 &  &  &  &  &  &  &  &  &  &  &  &  &  &  &  \\ 
&  A & .47 & .4 & .43 & .42 & .17 & -.04 & -.23 & -.06 & -.37 & .61 & .62 & .62 &  &  &  &  &  &  &  &  &  &  &  &  &  &  \\ 
&  $v^{(e)}$ & -.01 & -.05 & -.06 & .11 & .2 & .02 & -.04 & .1 & .2 & .44 & .4 & .4 & .19 &  &  &  &  &  &  &  &  &  &  &  &  &  \\ 
&  $v^{(f)}$ & .13 & .16 & .19 & .17 & .04 & .2 & .28 & .43 & .21 & .53 & .5 & .5 & .13 & .36 &  &  &  &  &  &  &  &  &  &  &  &  \\ 
&  $v^{(4)}$ & .31 & .34 & .35 & .35 & .03 & .24 & .33 & .44 & .12 & .6 & .62 & .62 & .25 & .31 & .85 &  &  &  &  &  &  &  &  &  &  &  \\ 
&  $v^{(8)}$ & .27 & .32 & .32 & .3 & .05 & .34 & .4 & .37 & .08 & .6 & .61 & .68 & .28 & .36 & .76 & .87 &  &  &  &  &  &  &  &  &  &  \\ 
&  $\tau$ & -.2 & -.17 & -.12 & -.26 & -.17 & .02 & .14 & .05 & .02 & -.7 & -.67 & -.64 & -.31 & -.52 & -.32 & -.29 & -.27 &  &  &  &  &  &  &  &  &  \\ 
  \hline
  \multirow{9}{*}{\rotatebox[origin=c]{90}{Stop Task}} &  $b^{(f)}$ & .5 & .51 & .46 & .29 & .34 & .48 & .35 & .27 & -.38 & .27 & .3 & .32 & .31 & -.02 & .19 & .26 & .27 & -.08 &  &  &  &  &  &  &  &  \\ 
&  $b^{(4)}$ & .52 & .53 & .48 & .31 & .33 & .48 & .36 & .27 & -.38 & .28 & .31 & .33 & .32 & -.01 & .19 & .26 & .29 & -.08 & 1 &  &  &  &  &  &  &  \\ 
&  $b^{(8)}$ & .53 & .55 & .5 & .32 & .32 & .48 & .36 & .27 & -.39 & .27 & .3 & .33 & .32 & -.03 & .18 & .26 & .28 & -.07 & .99 & .99 &  &  &  &  &  &  \\ 
&  A & .38 & .38 & .34 & .21 & .32 & .45 & .33 & .23 & -.33 & .14 & .18 & .2 & .25 & -.05 & .13 & .2 & .21 & .02 & .92 & .91 & .91 &  &  &  &  &  \\ 
&  $v^{(e)}$ & -.07 & -.09 & -.06 & -.05 & .02 & -.23 & -.24 & -.14 & .06 & .13 & .09 & .11 & .19 & .29 & -.04 & -.05 & .02 & -.14 & -.23 & -.21 & -.22 & -.29 &  &  &  &  \\ 
&  $v^{(f)}$ & .48 & .5 & .49 & .27 & .19 & .4 & .34 & .31 & -.36 & .32 & .32 & .34 & .29 & -.07 & .33 & .31 & .33 & -.09 & .79 & .79 & .78 & .63 & -.13 &  &  &  \\ 
 & $v^{(4)}$ & .52 & .55 & .53 & .3 & .2 & .43 & .37 & .31 & -.37 & .32 & .33 & .37 & .3 & -.05 & .31 & .31 & .35 & -.08 & .81 & .82 & .81 & .64 & -.1 & .96 &  &  \\ 
&  $v^{(8)}$ & .53 & .57 & .57 & .32 & .19 & .46 & .43 & .37 & -.36 & .28 & .3 & .33 & .3 & -.1 & .28 & .32 & .35 & 0 & .85 & .85 & .86 & .7 & -.14 & .94 & .96 &  \\ 
&  $\tau$ & -.16 & -.15 & -.1 & -.11 & -.14 & .01 & .05 & .06 & .11 & -.28 & -.27 & -.27 & -.15 & -.21 & -.04 & -.06 & -.06 & .46 & -.18 & -.19 & -.17 & 0 & -.3 & -.09 & -.13 & -.08 \\ 
\end{tabular}}
\label{tab:triplecor}
\end{table}
\end{landscape}

\end{document}